\documentclass[aps,prd, twocolumn ,
showkeys,
showpacs,
amssymb,
amsfonts,
epsf,
preprintnumbers,
nofootinbib,
superscriptaddress]{revtex4}

\usepackage[dvips]{graphicx}
\usepackage{bm,latexsym,amsmath,amssymb,amsfonts}
\usepackage{color}

\newcommand\beq{\begin{eqnarray}}
\newcommand\eeq{\end{eqnarray}}

\begin{document}

\title{Classification of cosmology with arbitrary matter in the Ho\v{r}ava-Lifshitz model}
\author{Masato Minamitsuji}
\affiliation{Center for Quantum Spacetime,
Sogang University, Shinsu-dong 1, Mapo-gu, 121-742 Seoul, South Korea}

\begin{abstract}
In this work,
we discuss the cosmological evolutions 
in the nonrelativistic and possibly renormalizable
gravitational theory, called the Ho\v{r}ava-Lifshitz (HL) theory.
We consider
the original HL model (type I), and
the modified version obtained by an analytic continuation
of parameters (type II).
We classify the possible cosmological evolutions
with arbitrary matter.
We will find a variety of cosmology.
\end{abstract}
\maketitle


\section{Introduction}

Studies on ultra-violet (UV) completion of gravity have a long history.
A lot of understanding has been obtained especially from string theory.
Very recently,
a new class of UV complete theory of gravity was
proposed by Ho\v{r}ava \cite{h2},
by generalizing the ideas discussed in Ref. \cite{h1}.
The theory does not have the full diffeomorphism
invariance as in Einstein's general relativity,
but only has a local Galilean invariance,
and Einstein's relativity may emerge
at the infra-red (IR) fixed point.
The model has been motivated by
a scalar field model discussed by Lifshitz \cite{lif},
to explain the quantum critical phenomena in condensed
matter physics,
in which the action has $z=2$,
where $z$ represents the {\it dynamical critical exponent}
given by
\beq
t\to \ell^z t,\quad
x^i\to \ell x^i\,. \label{scaling}
\eeq
The model proposed in Ref. \cite{h2} has the scaling dimension $z=3$
and is often called the Ho\v{r}ava-Lifshitz (HL) model.
%

There would be many implications of the HL model for cosmology,
which were firstly discussed in \cite{c,kk}.
In Ref. \cite{kk}, it was suggested
that
the divergence of speed of light in the UV region
may resolve the horizon problem.
The possibility of generation of the scale invariant spectrum
in the UV regime without inflation
was suggested in \cite {kk,mk,br,pi,cpt}.
Some detailed studies on perturbations in UV regime
have been presented in Refs. \cite{kk,mk,gao,chz,cpt}.
Tensor perturbations were investigated in Ref. \cite{ts},
which showed that primordial gravitational
waves are circularly polarized due to parity violation.
Phenomenological implications were suggested in \cite{mk2}.
The black hole solutions and related issues
have been studied very actively in \cite{lmp,mkim,cco,ks,kps}.
Some theoretical issues and generalizations of the HL model
have been discussed in \cite{kluson,nst,cys,svw,or}.

This article is devoted to
discuss the cosmological evolutions in HL model.
We consider the original HL model (type I),
and
the modified model obtained by an analytic continuation
of parameters (type II).
We classify the evolutions of the universe
with arbitrary kind of matter.
The article is constructed as follows.
In Sec. II, we briefly review the HL model
and give equations of motion in the cosmological background.
In Sec. III, we discuss the cosmological evolutions
in analogy with a point particle moving in a potential
in classical mechanics.
We classify the possible cosmological evolutions.
In Sec. IV, we give a brief summary before closing the article.

\section{Ho\v{r}ava-Lifshitz model and Cosmology}
\label{secII}


We briefly review the HL model.
The metric which has the local Galilean invariance can be written
in the form
\beq
&&ds^2=-N^2 dt^2
+g_{ij}\Big(dx^i+N^i dt\Big)
       \Big(dx^j+N^j dt\Big),
\eeq
and $N_{i}=g_{ij}N^j$, which is
similar to the ADM decomposition in
Einstein's general relativity \cite{adm}.
The scaling dimensions of various quantities
in the momentum unit are given by
$[t]=-3$,
$[x^i]=-1$,
$[N]=[g_{ij}]=0$
and $[N^i]=2$.
The dynamical variables are $N$, $N_i$ and $g_{ij}$,
which is very similar to the ADM decomposition in
Einstein's general relativity \cite{adm}.

The total action is composed of the kinetic part and potential one:
$S=S_K+S_V$.
The kinetic action is given by
\beq
&&S_K=\frac{2}{\kappa^2}
\int dt d^3 x\sqrt{g}N\Big(K_{ij}K^{ij}-\lambda K^2\Big),
\eeq
and $K^{ij}=g^{ik}g^{j\ell}K_{k\ell}$,
where the extrinsic curvature is defined by
\beq
K_{ij}=\frac{1}{2N}\Big(\frac{\partial}{\partial t}
g_{ij}-\nabla_i N_j-\nabla_j N_i\Big).
\eeq
$\nabla_i$ denotes the covariant derivative with respect to $g_{ij}$.
Correspondingly, with the detailed-balance condition,
the potential action is given as
\beq
S_V
&=&
\int dt d^3x\sqrt{g}N
\Big[
-\frac{\kappa^2}{2w^4}
 C_{ij}C^{ij}
\nonumber\\
&+&\frac{\kappa^2 \mu}{2w^2}\frac{\epsilon^{ijk}}{\sqrt{g}}
  R_{i\ell}\nabla_j R^{\ell}{}_{k}
-\frac{\kappa^2\mu^2}{8}R^{ij}R_{ij}
\nonumber\\
&+&\frac{\kappa^2\mu^2}{8(1-3\lambda)}
\Big(\frac{1-4\lambda}{4}R^2
   +\Lambda R-3\Lambda^2
\Big)
\Big].
\eeq
where $R_{ij}$ is Ricci tensor associated with $g_{ij}$,
and the Cotton tensor $C_{ij}$
is transverse, conserved and vanishing for all conformally flat spaces
and has scaling dimension $3$.
The full action is given by
\beq
S_I&=&S_K+S_V
\nonumber\\
&=&\int dt d^3x \sqrt{g}N
\Big[
\frac{2}{\kappa^2}
\Big(K_{ij}K^{ij}-\lambda K^2\Big)
-\frac{\kappa^2 }{2w^4}C_{ij}C^{ij}
\nonumber\\
&+&\frac{\kappa^2\mu}{2w^2}
\frac{\epsilon^{ijk}}{\sqrt{g}}
 R_{i\ell}\nabla_{j}R^{\ell}{}_k
-\frac{\kappa^2\mu^2}{8}R_{ij}R^{ij}
\nonumber\\
&+&\frac{\kappa^2\mu^2}{8(1-3\lambda)}
\Big(
 \frac{1-4\lambda}{4}R^2
+\Lambda R
-3\Lambda^2
\Big)
\Big],
\eeq
where scaling dimensions of parameters are given by
$[\kappa]=0$,
$[w]=0$,
$[\mu]=1$,
and $[\Lambda]=2$.
$\lambda$ represents a dynamical coupling constant,
susceptible to quantum corrections \cite{h2}.
For convenience,
one can rewrite the above action
\beq
S_I&=&\int dt d^3 x\sqrt{g}N
\Big[
\alpha\Big(K_{ij}K^{ij}-\lambda K^2\Big)
+\beta C_{ij}C^{ij}
\nonumber\\
&+&\gamma \frac{\epsilon^{ijk}}{\sqrt{g}}
 R_{i\ell}\nabla_j R^{\ell}{}_k
+\zeta R_{ij}R^{ij}
+\eta R^2
+\xi R
+\sigma
\Big]
\label{I}
\eeq
where
\beq
&&\alpha=\frac{2}{\kappa^2},\quad
\beta=-\frac{\kappa^2}{2w^4},\quad
\gamma=\frac{\kappa^2\mu}{2w^2},\quad
\zeta=-\frac{\kappa^2\mu^2}{8},
\nonumber\\
&&\eta=\frac{\kappa^2\mu^2}{8(1-3\lambda)}
\frac{1-4\lambda}{4},\quad
\xi=\frac{\kappa^2\mu^2}{8(1-3\lambda)}\Lambda,\quad
\nonumber\\
&&\sigma=\frac{\kappa^2\mu^2}{8(1-3\lambda)}
\big(-3\Lambda^2\big)\,.
\eeq
The action is invariant under the restricted class of diffeomorphisms
(foliation-preserving diffeomorphism)
$t'=h(t)$ and
$\big(x'\big)^i=h^i(t,x^i)
$.
For $\lambda=1/3$,
the theory has classical, anisotropic conformal invariance.
Comparing the action with that of general relativity in ADM formalism,
one can read the emergent constants
\beq
c=\frac{\kappa^2\mu}{4}\sqrt{\frac{\Lambda}{1-3\lambda}},
\quad
G=\frac{\kappa^2}{32\pi c},
\quad
\Lambda_E=\frac{3}{2}\Lambda\,,
\label{emergent}
\eeq
appearing in IR.
Einstein's general relativity is recovered for $\lambda=1$.


Following Ref. \cite{lmp},
one can make an analytic continuation of parameters
$\mu\to i \mu$
and
$w^2\to -i w^2$.
Then,
the above action can be rewritten as
\beq
S_{II}&=&\int dt d^3 x\sqrt{g}N
\Big[
\alpha_2\Big(K_{ij}K^{ij}-\lambda K^2\Big)
+\beta_2 C_{ij}C^{ij}
\nonumber\\
&+&\gamma_2 \frac{\epsilon^{ijk}}{\sqrt{g}}
 R_{i\ell}\nabla_j R^{\ell}{}_k
+\zeta_2 R_{ij}R^{ij}
+\eta_2 R^2
+\xi_2 R
+\sigma_2
\Big]
\label{II}
\eeq
where new parameters are defined by
$\alpha_2=\alpha$,
$\beta_2=-\beta$,
$\gamma_2=-\gamma$,
$\zeta_2=-\zeta$,
$\eta_2=-\eta$,
$\xi_2= -\xi$
and
$\sigma_2=-\sigma$.
The emergent speed of light is given by
$
c=(\kappa^2 \mu/4)\sqrt{\Lambda/(3\lambda-1)}
$.
For convenience, in this article,
we call the model Eq. (\ref{I}) "type I"
and
the model Eq. (\ref{II}) "type II".


\vspace{0.2cm}

For the cosmological background,
i.e., the homogeneous and isotropic background,
$N= N(t)$, $N_i= 0$ and $g_{ij}=a^2(t) \gamma_{ij}$,
where $\gamma_{ij}$ is maximally symmetric 3-space whose curvature is given
by $R^{\gamma}_{ij}=2k \gamma_{ij}$ and $R^{\gamma}=6k$.
In the cosmological spacetime,
the Cotton tensor vanishes
and the equations of motion get simplified.

In the type I theory,
the equations of motion (Friedmann equations) are given by
\beq
&&H^2+\frac{\kappa^4\mu^2(k-a^2\Lambda)^2}{16(1-3\lambda)^2 a^4}
=\frac{\rho}{3\alpha(3\lambda-1)},
\nonumber\\
&&
2\Big(\dot{H}+\frac{2}{3}H^2\Big)
-\frac{\kappa^4 \mu^2}{16a^4(1-3\lambda)^2}
\big(k+3a^2\Lambda\big)\big(k-a^2\Lambda\big)
\nonumber\\
&=&-\frac{p}{\alpha(3\lambda-1)},
\label{youko}
\eeq
where $H:=\dot{a}/(aN)$ (in the later discussion, we simply set $N=1$)
and
$\rho$ and $p$ are the energy density and pressure
of the matter sector.
In the type II theory,
by changing $\mu^2\to -\mu^2$, the equations of motion become
\beq
&&H^2-\frac{\kappa^4\mu^2(k-a^2\Lambda)^2}{16(1-3\lambda)^2 a^4}
=\frac{\rho}{3\alpha(3\lambda-1)},
\nonumber\\
&&
2\Big(\dot{H}+\frac{2}{3}H^2\Big)
+\frac{\kappa^4 \mu^2}{16a^4(1-3\lambda)^2}
\big(k+3a^2\Lambda\big)\big(k-a^2\Lambda\big)
\nonumber\\
&=&-\frac{p}{\alpha(3\lambda-1)}\,.
\label{mika}
\eeq
It is convenient to represent the Friedmann equation
in analogy with a dynamics of a point particle in a potential
in the classical mechanics,
$\dot{a}^2+V_m(a)=0$
($m=I, II$),
where
\beq
&&V_m(a)=\epsilon_m \frac{\kappa^4\mu^2(k-a^2\Lambda)^2}{16(1-3\lambda)^2 a^2}
      -\frac{\rho a^2}{3\alpha(3\lambda-1)},
\label{pot}
\eeq
where $\epsilon_I=+ 1$ and $\epsilon_{II}=-1$, respectively.
In the absence of the matter $\rho=0$,
the effective potential is not sensitive
to the value of $\lambda$ unless $\lambda=1/3$.
With matter,
it depends on
whether $\lambda$ is larger or smaller than $1/3$.
In the regime $\lambda>1/3$, since $\alpha>0$,
the sign of the energy density terms in Eq. (\ref{pot})
is positive.
While, in the regime where $\lambda <1/3$,
the sign of the same term is flipped.
In the case $\lambda=1/3$, where the theory develops an anisotropic
conformal invariance, the scale factor becomes non-dynamical
as seen in Eq. (\ref{youko}) and (\ref{mika}).
In the later discussions, we assume $\lambda\neq 1/3$.
Note that the case of $\lambda<1/3$ induces
the repulsive gravitational force (see \eqref{pot}), and also
the perturbation about the flat background provides
a ghost-like scalar mode \cite{h2},
which is potentially dangerous if it is coupled to the matter.
Thus, this branch may not be realistic.
But for the mathematical completeness, in this article, 
we will include the case of $\lambda<1/3$ into our analysis.

\section{Cosmological evolutions}

In this section, we discuss the classification
of the cosmological evolutions with matter in the HL model.
We assume that the energy density of the matter is nonnegative
$\rho\geq 0$, which can be parameterized as
$\rho=\rho_0 a_0^n/a^n$,
where $n$ is arbitrary integer and $\rho_0\geq 0$.
One might think that it would be enough to 
consider the matter of $n=6$ with the equation of state $p=\rho$,
because in the nonrelativistic theory with the dynamical critical exponent $z$,
the equation of state of the massless particles becomes $p=(z/3)\rho$.
But it seems to be too restrictive.
The HL theory itself is purely a gravitational theory
and does not specify the nature of the matter.
In addition, there is no particular reason
why the matter sector may not share the same dynamical 
critical exponent with the gravity sector,
although it may be plausible.
Therefore, for completeness of our analysis, 
it is appropriate to keep $n$ to be an arbitrary integer.

It is important to note two important properties.
In the type I model and in the presence of the matter $\rho>0$
there is no solution for $\lambda<1/3$, because
the potential $V_{I}(a)$ defined by \eqref{pot} is always positive.
In the type II model and in the presence of the matter $\rho>0$
there are always monotonic solutions
\footnote{By "monotonic", we mean the universe
which expands from $a=0$ to $a=\infty$
and contracts from $a=\infty$ to $a=0$,
without any collapse or bounce.
}
,
for $\lambda>1/3$, because
the potential $V_{II}(a)$ defined by \eqref{pot}
is always negative.
Thus, in the presence of the matter,
we will basically focus
on the case of $\lambda>1/3$ for the type I model
and
on the case of $\lambda<1/3$ for the type II model.

In the simplest case $k=\Lambda=0$,
both in type I and II theories,
the solutions are given by
\beq
a=\left(\frac{\rho_0 n^2 a_0^n}{12\alpha(3\lambda-1)}\right)^{1/n}
t^{2/n},
\eeq
which is only possible for $\lambda>1/3$.

\subsection{The vacuum case}

Firstly,
we consider the case of a vacuum solution $\rho=0$.

\paragraph{Type I model}

For the nonzero $k$ and $\Lambda$,
there is only the static solution for $k/\Lambda >0$,
which is given by
$
a_I=\sqrt{k/\Lambda}.
$
where the effective potential Eq. (\ref{pot}) is vanishing.

For $k=0$ (but $\Lambda\neq 0$)
or
for $\Lambda=0$ (but $k=0$), the potential
becomes positive everywhere
and there is no cosmological solution.

\paragraph{Type II model:}
For $k \neq 0$ and $k/\Lambda>0$,
there is an exact solution given by
\beq
a_{II}(t)=\sqrt{\frac{k}{\Lambda}}
         \sqrt{1+c_1 e^{\pm 2H_0t}}, \quad
H_0:=\Big|\frac{\kappa^2 \mu \Lambda}{4(3\lambda-1)}\Big|\,,
\label{acc}
\eeq
where $c_1$ is an arbitrary constant, which can be positive or negative.
There are four possibilities.
The first one is that
the universe starts from $a=0$ and ever approaches $a=\sqrt{k/\Lambda}$.
The second possibility is
that the universe starts to contract from the infinity
and ever approaches $a=\sqrt{k/\Lambda}$.
The third and fourth ones are oppositely
that it starts from $a=\sqrt{k/\Lambda}$,
and then
collapses at $a=0$
or
ever expands toward $a\to\infty$,
respectively.

For $k=0$ (but $\Lambda \neq 0$), there is the de Sitter solution
\beq
a_{II}(t)=a_0 e^{\pm H_0 t},\quad
H_0=\Big|\frac{\kappa^2 \mu \Lambda}{4(3\lambda-1)}\Big|.
\eeq
For $\Lambda=0$ (but $k\neq 0$), the universe is just like
the radiation-dominated one in the relativistic theory,
where the solution is given by
\beq
a_{II}(t)=\left|\frac{\kappa^2\mu k}{2(3\lambda-1)}\right|^{1/2}t^{1/2}.
\eeq


\subsection{The case $n\geq 5$}

Then, we study the cosmology with matter.
In this section, we discuss the case $n\geq 5$.
The radiation energy density would scale as $a^{-6}$
in the UV regime, not as $a^{-4}$ in the relativistic cosmology,
because
in the UV regime
the dispersion relation of a massless particle becomes
$\omega \propto p^3$,
where $p$ is the physical momentum and is redshifted as $p\propto 1/a$,
and the radiation energy density
is given by $\rho\sim n \omega \propto 1/a^6$, where
$n\propto 1/a^3$ is number density of massless particles.

\vspace{0.2cm}

\paragraph*{Type I model ($\lambda>1/3$):}
For all the choices of $\Lambda$ and $k$,
big crunch solutions are obtained.

\paragraph*{Type II model ($\lambda<1/3$):}
For all the choices of $\Lambda$ and $k$,
bouncing solutions are obtained.


\subsection{The case $n=4$}

This case corresponds to radiation contribution
in IR regime.

\paragraph*{Type I model ($\lambda>1/3$):}
If
\beq
\frac{\kappa^4 \mu^2 k^2}{16(3\lambda-1)}>\frac{\rho_0 a_0^4}{3\alpha}
\label{case1}
\eeq
there is no solution, or there are cyclic universes.
If
\beq
\frac{\kappa^4 \mu^2 k^2}{16(3\lambda-1)}<\frac{\rho_0 a_0^4}{3\alpha},
\label{case2}
\eeq
there are big crunch solutions.

For $k=0$ (but $\Lambda\neq 0$),
which is the special case of Eq. (\ref{case2}),
there are big crunch solutions.
For $\Lambda=0$ (but $k \neq 0$), there is no solution for the case Eq. (\ref{case1})
and there are monotonic solutions for the case Eq. (\ref{case2}).

The exact solution is given by
\beq
a_I(t)=\frac{1}{\sqrt{2}H_0}
 \sqrt{K-(K^2-4C H_0^2)^{1/2}\cos(2H(t-t_0))}
\label{cp1}
\eeq
where
\beq
&&C=\frac{\kappa^4 \mu^2k^2}{16(1-3\lambda)^2}
+\frac{\rho_0a_0^4}{3\alpha(1-3\lambda)}\,,
\nonumber\\
&&
H_0^2:=\frac{\kappa^4\mu^2\Lambda^2}{16(1-3\lambda)^2},
\quad
K:=\frac{\kappa^4\mu^2 k\Lambda}{8(1-3\lambda)^2}\,.
\eeq
The solution only exists for $K^2>4CH_0^2$,
which is consistent with $\lambda>1/3$.

\paragraph*{Type II model ($\lambda<1/3$):}
If
\beq
\frac{\kappa^4 \mu^2 k^2}{16(1-3\lambda)}<\frac{\rho_0 a_0^4}{3\alpha}
\label{case3}
\eeq
there are bouncing solutions.
If
\beq
\frac{\kappa^4 \mu^2 k^2}{16(1-3\lambda)}>\frac{\rho_0 a_0^4}{3\alpha}
\label{case4}
\eeq
there are monotonic solutions.

For $k=0$ (but $\Lambda\neq 0$), which is just the special case of Eq. (\ref{case3}),
there are bouncing solutions.
For $\Lambda=0$ (but $k\neq 0$), there is no solution for Eq. (\ref{case3})
and there are monotonic solutions for Eq. (\ref{case4}).

The exact solution is given by
\beq
a_{II}(t)=\frac{1}{\sqrt{2}H_0}
 \sqrt{K\pm (K^2-4C H_0^2)^{1/2}\cosh(2H(t-t_0))},
\label{cp2}
\eeq
for $K^2>4C H_0^2$,
and
\beq
a_{II}(t)=\frac{1}{\sqrt{2}H_0}
 \sqrt{K\pm (4C H_0^2-K^2)^{1/2}\sinh(2H(t-t_0))},
\label{cp3}
\eeq
for $K^2<4 C H_0^2$, where
\beq
&&C=\frac{\kappa^4 \mu^2k^2}{16(1-3\lambda)^2}
+\frac{\rho_0a_0^4}{3\alpha(3\lambda-1)}\,,
\nonumber\\
&&
H_0^2:=\frac{\kappa^4\mu^2\Lambda^2}{16(1-3\lambda)^2},
\quad
K:=\frac{\kappa^4\mu^2 k\Lambda}{8(1-3\lambda)^2}\,.
\eeq
The first and second solution corresponds
to $\lambda<1/3$ (bouncing solutions)
and
$\lambda>1/3$ (monotonic solutions).


\subsection{The case $n=3$}

The case $n=3$ corresponds to that of massive particles:
$\rho\sim m n\propto 1/a^3 $, where $n$ is the number density of
particles with mass $m$.

\paragraph*{Type I model ($\lambda>1/3$):}
In the type I model,
there is no cosmological solution unless $\lambda>1/3$.
For $\lambda<1/3$
In this case,
there is no solution or a cyclic universe is available.
For $k=0$  (but $\Lambda \neq 0$), big crunch solutions are obtained, while
for $\Lambda=0$  (but $k \neq 0$) bouncing universes are.

\paragraph*{Type II model:}

For $\lambda<1/3$,
there are monotonic,
or big crunch, or bouncing solutions.
For $k=0$ (but $\Lambda \neq 0$),
which just the special case of Eq. (\ref{case3}),
there are bouncing solutions.
For the case of $\Lambda=0$ (but $k \neq 0$),
there are big crunch solutions.

For $\lambda>1/3$ and $k=0$ (but $\Lambda \neq 0$),
there is the exact solution
\beq
a_{II}(t)
&=&\Big\{
\frac{4}{\kappa^2\mu\Lambda}
\sqrt{\frac{\kappa^2(3\lambda-1)\rho_{0}a_0^3}{6}}
\nonumber\\
&\times&
\sinh
\Big[
\frac{3\kappa^2\mu\Lambda}{8(3\lambda-1)} (t-t_0)
\Big]
\Big\}^{2/3}.
\eeq


\subsection{The case $n=1$ and $n=2$}

\paragraph*{Type I model ($\lambda>1/3$):}

For $\lambda>1/3$,
in this case,
there is no solution, or
there are cyclic solutions.
For $k=0$ (but $\Lambda\neq 0$), there are big crunch solutions.
For $\Lambda=0$ (but $k\neq 0$), there are bouncing solutions.

\paragraph*{Type II model ($\lambda<1/3$):}

There are monotonic, or big crunch,
or bouncing solutions.
For $k=0$ (but $\Lambda\neq 0$), there are bouncing solutions.
For $\Lambda=0$ (but $k\neq 0$), there are big crunch solutions.


\subsection{The case $n=0$}

\paragraph*{Type I model ($\lambda>1/3$):}

For $\Lambda\neq 0$ and $k\neq 0$,
if
\beq
\frac{\kappa^4\mu^2\Lambda^2}{16(3\lambda-1)}
>\frac{\rho_0}{3\alpha}, \label{kink1}
\eeq
there is no solution,
or there are cyclic solutions.
If
\beq
\frac{\kappa^4\mu^2\Lambda^2}{16(3\lambda-1)}
<\frac{\rho_0}{3\alpha},\label{kink2}
\eeq
there are bouncing solutions.

For $k=0$ (but $\Lambda\neq 0$),
if Eq. (\ref{kink1}) there is no solution,
and
if Eq. (\ref{kink2}) there are monotonic solutions.
For $\Lambda=0$ (but $k \neq 0$), there are
bouncing solutions.

The exact solution is given by
Eq. (\ref{cp1})
where
\beq
&&C=\frac{\kappa^4 \mu^2k^2}{16(1-3\lambda)^2},\quad
K:=\frac{\kappa^4\mu^2 k\Lambda}{8(1-3\lambda)^2}\,,
\nonumber\\
&&
H_0^2:=\frac{\kappa^4\mu^2\Lambda^2}{16(1-3\lambda)^2}
+\frac{\rho_0}{3\alpha(1-3\lambda)}\,.
\eeq
The solution only exists for $K^2>4CH_0^2$,
which is consistent with $\lambda>1/3$.

\paragraph*{Type II model ($\lambda<1/3$):}

For $\Lambda\neq 0$ and $k\neq 0$,
if
\beq
\frac{\kappa^4\mu^2\Lambda^2}{16(1-3\lambda)}
>\frac{\rho_0}{3\alpha}, \label{kink3}
\eeq
there are monotonic, or big crunch or bouncing solutions.
If
\beq
\frac{\kappa^4\mu^2\Lambda^2}{16(1-3\lambda)}
<\frac{\rho_0}{3\alpha},\label{kink4}
\eeq
there are big crunch solutions.

For $k=0$ (but $\Lambda\neq 0$),
if Eq. (\ref{kink3}) there are monotonic solutions,
and
if Eq. (\ref{kink2}) there is no solution.
For $\Lambda=0$ (but $k \neq 0$), there are
big crunch solutions.

The exact solution is given by
Eq. (\ref{cp2})
for $K^2>4C H_0^2$,
and
Eq. (\ref{cp3})
for $K^2<4 C H_0^2$, where
\beq
&&C=\frac{\kappa^4 \mu^2k^2}{16(1-3\lambda)^2},
\quad
H_0^2:=\frac{\kappa^4\mu^2\Lambda^2}{16(1-3\lambda)^2}
+\frac{\rho_0}{3\alpha(1-3\lambda)}\,,
\nonumber\\
&&
K:=\frac{\kappa^4\mu^2 k\Lambda}{8(1-3\lambda)^2}
\,.
\eeq
The first and second solution corresponds
to $\lambda<1/3$ 
and
$\lambda>1/3$.


\subsection{The case $n\leq -1$}


\paragraph*{Type I model ($\lambda>1/3$):}
For all the choices of $\Lambda$ and $k$,
there are bouncing solutions.

\paragraph*{Type II model ($\lambda<1/3$):}
For all the choices of $\Lambda$ and $k$,
there are big crunch solutions

\section{Summary}

We investigated evolutions of homogeneous and
isotropic universe in the recently proposed nonrelativistic,
renormalizable gravitational theory with an anisotropic scaling $z=3$
(called the Ho\v{r}ava-Lifshitz (HL) theory),
where $z$ is the dynamical critical exponent defined in Eq. (\ref{scaling}).
We considered the original theory (type I)
and the modified model obtained by an analytic continuations of
parameters in the original theory (type II).
The dimensionless, dynamical coupling parameter
$\lambda$ is contained into the theory,
which is sensible
to the quantum corrections.
The theory has IR fixed point and for $\lambda=1$
the theory coincides with general relativity in IR regime.
For $\lambda=1/3$, the theory has an anisotropic conformal
invariance and scale factor becomes non-dynamical.
In this article, we
did not restrict the range of $\lambda$.
We also have assumed
that the matter energy density is positive $\rho>0$.

In the type I model,
we found that
\begin{itemize}

\item
For the vacuum universe,
there is a static solution, $a=\sqrt{k/\Lambda}$,
for $\Lambda\neq 0$ and $k/\Lambda>0$ .

\item
With arbitrary kind of matter,
there is no cosmological solution for $\lambda<1/3$

\item
For $\Lambda=0$ (but $k\neq 0$)
and with matter ($\lambda>1/3$),
there are bouncing solutions for $n\leq 3$ \cite{br},
and
there are big crunch solutions for $n\geq 5$.
The case $n=4$ is marginal:
If Eq. (\ref{case1}) there is no solution,
but if Eq. (\ref{case2}),
there are monotonic solutions.

\item
For $k=0$ (but $\Lambda\neq 0$)
and with matter ($\lambda>1/3$),
there are bouncing solutions for $n\leq -1$,
and
there are big crunch solutions for $n\geq 1$.
The case $n=0$ is marginal:
If Eq. (\ref{kink1}) there is no solution,
but if Eq. (\ref{kink2}),
there are monotonic solutions.

\end{itemize}

In the type II model,
we found that
\begin{itemize}

\item
For $\Lambda\neq 0$ (and $k/\Lambda>0$) and in the absence of matter,
there are exact solutions given by Eq. (\ref{acc}) \cite{lmp}.
They are solutions approaching $a=\sqrt{k/\Lambda}$
or starting from $a=\sqrt{k/\Lambda}$.

\item
With arbitrary kind of matter ($\lambda>1/3$),
there are monotonic solutions for any $n$.

\item
For $\Lambda=0$ (but $k\neq 0$)
and with matter ($\lambda<1/3$),
there are big crunch solutions for $n\leq 3$ \cite{br},
and
there are bouncing solutions for $n\geq 5$.
The case $n=4$ is marginal:
If Eq. (\ref{case3}) there is no solution,
but if Eq. (\ref{case4}),
there are monotonic solutions.

\item
For $k=0$ (but $\Lambda\neq 0$)
and with matter ($\lambda<1/3$),
there are big crunch solutions for $n\leq -1$,
and
there are bouncing solutions for $n\geq 1$.
The case $n=0$ is marginal:
If Eq. (\ref{kink4}) there is no solution,
but if Eq. (\ref{kink3}),
there are monotonic solutions.

\end{itemize}

It may be helpful to clarify the characteristic differences of the cosmological evolutions in the HL theory from those in general relativity.

\begin{itemize}

\item
In the case of the type I theory:
supposing that $\rho > 0$ and
$\lambda>1/3$ (the gravity is attractive),
in the case of $k=0$ and $\Lambda=0$, 
there is no essential difference
from the case of general relativity.
In the case of nonzero $k$,
the particular difference from the case of general relativity
is the existence of the bouncing evolutions 
in the absence of matter which breaks the weak energy condition.
Note that
in general relativity, it is impossible to realize
the bouncing Universe under the weak energy condition.
In the HL theory, the corrections due to the 
higher spatial curvature terms could effectively break it.
Therefore, this is the novel effect in the HL gravity.
On the other hand,
even in the flat ($k=0$) or open ($k<0$) Universe, with matter of $n\geq 0$,
the presence of non-zero $\Lambda$ induces the maximal size of 
the Universe, i.e., the big crunch solutions.

\item  
In the case of the type II theory:
supposing that $\rho > 0$ and
$\lambda>1/3$ (the gravity is attractive),
in the case of $k=0$ and $\Lambda=0$,  
there is no essential difference
from the case of general relativity.
However, in all the cases,
the effective potential Eq. (\ref{pot}) is always non-positive.
Thus, even in the case of the closed ($k>0$) Universe, 
there is no recollapsing Universe,
which is the crucial difference from the case of general relativity.

\end{itemize}

In this article, 
for simplicity,
the detailed balance condition was assumed,
but it may not be essential.
The absence of the detailed balance condition introduces
an additional $a^{-4}$ contribution to the effective potential 
Eq. (\ref{pot}) (See e.g.,\cite{svw}), and leads to
a richer variety of the cosmological evolutions. 
In addition,
in the type I theory with the detailed balance condition,
to obtain the attractive gravitational force, one has to choose
$\lambda>1/3$ (see Eq. (\ref{pot})).
Then in order to have the real emergent speed of light,
from Eq. (\ref{emergent}) the condition of $\Lambda<0$ has to be imposed.
Namely there is only the anti-de Sitter vacuum ($\Lambda_E<0$).
Although this is not the case for the type II theory, 
the detailed balance may not be suitalbe to obtain the realistic cosmology. 
For the detailed classifications in the case without
the detailed balance condition, see \cite{min}.

Before closing this article, we shall note some properties of the model,
which would be important for future studies.
We mainly focused on the possible cosmological
evolutions in the UV regime
and ignored the process of the recovery of the cosmology
in Einstein's relativity.
The UV cosmology should be smoothly matched to
that of IR.
But, there still seem to be some unclear points about this connection.
First,
we assumed that $\lambda$ is a constant
but actually
$\lambda$ would be sensitive to the perturbative corrections.
$\lambda$ would eventually approach some IR fixed point
and variation of $\lambda$ would affect the cosmology.
Second,
in Ref. \cite{h2,h1,ks,chz,cpt},
it has been pointed out that in the HL
theory there is another (scalar) degree of freedom
other than those appearing in general relativity,
which corresponds to a scalar mode in the linearized theory.
The analysis of perturbations about a flat spacetime indicates that
the kinetic term of this scalar mode becomes a ghost
for $\lambda<1/3$ and $\lambda\geq 1$,
and may be important for stability of the solution,
if this mode is coupled to the matter perturbations \cite{h2,h1,ks}.
It also has been pointed out that in the cosmological background
this mode may be useful to produce the scale invariant spectrum
without inflaton in the UV regime, if the detailed balance
condition is broken \cite{cpt}.
The decoupling of scalar mode in approaching IR
would also be essential to recover the relativity.


\section*{Acknowledgements}
This work was supported by the Korea Science and Engineering Foundation (KOSEF) grant No. R11-2005-021, founded by the Korean Government (MEST) through the Center for Quantum Spacetime (CQUeST) of Sogang University.

\end{document}